\documentstyle[sprocl,psfig]{article}

\def\Journal#1#2#3#4{{#1} {\bf #2}, #3 (#4)}

\def\NPB{{\em Nucl. Phys.} B}
\def\PLB{{\em Phys. Lett.}  B}
\def\PRL{\em Phys. Rev. Lett.}
\def\PRD{{\em Phys. Rev.} D}

\def\be{\begin{equation}}
\def\ee{\end{equation}}
\def\bea{\begin{eqnarray}}
\def\eea{\end{eqnarray}}

\newcommand{\lsim}%
{\mathrel{\mbox{\raisebox{-1.0ex}%
{$\stackrel{\textstyle <}{\textstyle \sim}$}}}}

\newcommand{\preprintnumber}[1]
{\begin{flushright}
  \begin{tabular}{l} #1 \end{tabular}
  \end{flushright}}

\begin{document}

\preprintnumber{
KEK-TH-684\\
hep-ph/0004025
}
\vspace{3em}

\title{$\mu \rightarrow e \gamma$ AND $\mu \rightarrow 3e$ PROCESSES WITH
 POLARIZED MUON AND SUPERSYMMETRIC GRAND UNIFIED THEORIES \footnote{
Talk given at the Workshop on High Intensity Muon Sources (HIMUS99),
 Tsukuba, Japan, 1-4 Dec 1999.}}

\author{Ken-ichi Okumura}

\address{Theory Group, KEK, Tsukuba, Ibaraki, 305-0801 Japan\\
 and \\
Department of Accelerator Science, The Graduate University for 
Advanced Studies,\\
 Tsukuba, Ibaraki, 305-0801 Japan, 
\\E-mail: ken-ichi.okumura@kek.jp}


\maketitle\abstracts{Lepton flavor violating processes
 $\mu \rightarrow e \gamma$
 and $\mu \rightarrow 3e$ with polarized muons are studied
 in the supersymmetric grand unified theory (SUSY GUT).
 As a result of a detailed numerical calculation, 
 it is shown that the P- and T-odd asymmetries
 defined with the help of
 the muon polarization and the ratio of two branching fractions
 make a good contrast between the SU(5) and SO(10) SUSY GUT.
 These observables are useful to extract differences of the two theories.
 In particular, the P-odd asymmetry of $\mu \rightarrow e \gamma$
  varies $100\%$-$-100\%$ in SO(10) whereas it is $100\%$ in
 SU(5) and the T-odd asymmetry of $\mu \rightarrow 3e$ can
 reach $15\%$ in SU(5) within the EDM constraints whereas it is
 small in SO(10).} 

\section{Introduction}
%
In the standard model (SM) lepton flavor is conserved because
 the matter contents of the SM and the gauge symmetry forbid
 lepton flavor violating renormalizable couplings. 
However, matter contents beyond the SM can easily accommodate
 violation of lepton flavor.
In the supersymmetric (SUSY) extension of the SM,
scalar partners of the ordinary leptons, which are introduced
to solve the gauge hierarchy problem, become a source of
 lepton flavor violation (LFV).
 Terms in the SUSY Lagrangian which
 violate SUSY without the quadratic divergence
 (the soft SUSY breaking terms) are not necessarily diagonal with 
 respect to the flavor indices and become a source of LFV. 
 Even if their origin is flavor-blind at a very high energy scale
 as in the case of the minimal supergravity model (minimal SUGRA),
 radiative corrections from LFV interactions between that scale
 and the electro-weak (EW) scale can induce LFV off-diagonal elements
 in the slepton mass matrices \cite{hkr86}.
At a low energy scale, an existence of such interactions can appear
 as LFV processes such as $\mu \rightarrow e \gamma$ and
 $\mu \rightarrow 3e$ through loop diagrams including SUSY partners
 suppressed only by the power of their mass scale.
 Then, the precise measurement of such a low energy process is
 a sensitive probe for the LFV
 interactions at a very high energy scale.
In particular, grand unified theory (GUT) predicts unification of 
 quarks and leptons and the flavor mixing in quark sector means
 that GUT interactions violate lepton flavor conservation.
The pattern of the LFV interactions reflects
 the structure of SUSY GUT.
LFV processes which are induced from such interactions
 can be just below the current experimental bounds
 \cite{bh94,bhs95}.
There are experimental projects which aim to
 explore the region of more than three orders of magnitude
 below the current limits \cite{ex}.
In this article we discuss the possibility to use various P- and T- odd
 asymmetries in $\mu \rightarrow e\gamma$ and $\mu \rightarrow 3e$ 
 with polarized muons to study
 the LFV interactions above the GUT scale
 and distinguish different SUSY models: SU(5) and SO(10) SUSY GUT
 \footnote{This talk is based on the work in reference \cite{K.O.}.}.

\section{$\mu \rightarrow e \gamma$ and $\mu \rightarrow 3e$ processes in SU(5) and SO(10) SUSY GUT} 
First, we introduce the minimal SU(5) SUSY GUT\cite{su5} and discuss
 the qualitative feature of LFV in this theory.
In the SU(5) SUSY GUT, all the matter fields in the minimal
 supersymmetric standard model (MSSM) are embedded in three
 generations of {\bf 10} representation
 $T_i(Q_i,U^c_i,E^c_i)$ and {\bf \=5} representation
 $\overline{F}_i(D^c_i,L_i)$ of SU(5). 
Two Higgs doublets are embedded in {\bf 5} ($H(H_C,H_2)$)and {\bf \=5}
 ($\overline{H}(\overline{H}_C,H_1)$) representations with
 newly introduced colored Higgs fields $H_C$ and $\overline{H}_C$.
 The renormalizable Yukawa superpotential is written as follows.
\begin{eqnarray}
{\cal W}_{SU(5)} = \frac{1}{8}\epsilon_{abcde}(\hat{y}_{u})_{i}
                   T_{i}^{ab}T_{i}^{cd}H^e 
             + (y_{d})_{ij}\overline{F}_{ia}T_{j}^{ab}\overline{H}_b,
\label{eq:Lagrangian su5}
\end{eqnarray}
where we choose the basis in which the up-type Yukawa coupling constant
 $y_u$ is diagonal.
The important point in this formula for LFV
 is that the right-handed leptons have GUT
 interaction through up-type Yukawa coupling
 constant: $(\hat{y}_u)_{i} U^c_i E^c_i H_C$.
 Because of this GUT interaction including the large top
 Yukawa coupling constant, even if the soft SUSY 
breaking terms have the minimal-SUGRA-type universal
 structure at the Planck scale,  
radiative corrections between the Planck scale and the GUT scale
 reduces the third generation 
slepton mass compared to the first and
 the second generation.
The mass difference can be approximated as follows:
\begin{eqnarray}
\Delta m^{2} &\simeq& \frac{3}{8\pi^{2}}|(\hat{y}_{u})_{3}|^{2}m_{0}^{2}
(3 + |A_{0}|^{2})\ln(\frac{M_{P}}{M_{G}}). 
\end{eqnarray}
In order to discuss the LFV processes,
 it is convenient to go to the basis where the lepton 
Yukawa coupling constant is diagonal.
\begin{eqnarray}
 V_{R}y_{e}V_{L}^{\dag} &=& diagonal
\end{eqnarray}
where $V_{R}$ and $V_{L}$ are unitary matrices.
In this basis, the right-handed sleptons have LFV
 off-diagonal elements in the mass matrix as follows:
\begin{eqnarray}
(m_{E}^{2})_{ij} \simeq -(V_R^T)_{3i}(V_R^{\dag})_{3j} \Delta m^{2}.
\label{eq:m_R su5} 
\end{eqnarray}
A similar formula can be obtained for the left-right mixing mass matrix
 which is induced from the trilinear scalar coupling in the soft SUSY breaking terms.
All the amplitudes of $\mu^+ \rightarrow e^+ \gamma$ and
 $\mu^+ \rightarrow e^+e^+e^-$ processes pick up these off diagonal
 elements and they are proportional to the combination of
 unitary matrix elements $\lambda_{\tau}=(V_R^T)_{32}(V_{R}^{\dag})_{31}$.
In the minimal model, $V_{R}$ is written by 
the transposed of the CKM matrix because of the GUT
relation $y_e = y_d^T$. 
However, this relation can not explain the ratio of the 
first and second generation quark and lepton masses.
In the realistic model, higher dimensional operators
at the GUT scale can explain such a mismatch.
With these new operators, $V_R$ can be different from
 the corresponding CKM matrix element and the branching ratios
 themselves have considerable model dependence.
Here, instead of dealing with a detail of models
 we focus on the 
observables which are expressed by the 
ratio of amplitudes and insensitive to the value of $\lambda_{\tau}$.
 Next we explain the case of the minimal SO(10) SUSY GUT
\cite{so10}.
 In this theory, all the matter fields in the MSSM are
 unified in three generations of 16 representation of SO(10) ($\Psi_i$). 
We introduce two 10 representation Higgs fields $\Phi_u$, $\Phi_d$
for the up- and down-type Yukawa couplings to reproduce the CKM matrix.
The renormalizable Yukawa superpotential can be written
 as follows\cite{dh95}:
\begin{eqnarray}
\label{eq:Lagrangian so10}
 {\cal W}_{ SO(10)} &=& \frac{1}{2}(\hat{y}_{u})_{i}\Psi_{i}\Phi_{u}\Psi_{i} 
             + \frac{1}{2}(y_{d})_{ij}\Psi_{i}\Phi_{d}\Psi_{j}.
\end{eqnarray}
The important difference from the SU(5) case is 
that the left-handed leptons also have GUT interaction
 through the up-type Yukawa coupling constant.
Because of this GUT interaction the left-handed sleptons
also have LFV off-diagonal elements in the mass matrix $(m_L)_{ij}$
even if we have the universal scalar mass at the Planck scale.

The LFV off-diagonal elements of the slepton mass matrix discussed
 above induce $\mu \rightarrow e\gamma$
 and $\mu \rightarrow 3e$ through loop diagrams which include
 sleptons, neutralino and chargino.
The most general form of the effective Lagrangian which 
describes these processes can be parameterized using
 the Lorentz invariance, the gauge invariance
and the Fierz rearrangement as follows: 
\begin{eqnarray}
\label{eq:effective_Lagrangian}
{\cal L} &=& -\frac{4G_F}{\sqrt{2}}\{  
        {m_{\mu }}{A_R}\overline{\mu_{R}}
        {{\sigma }^{\mu \nu}{e_L}{F_{\mu \nu}}}
       + {m_{\mu }}{A_L}\overline{\mu_{L}}
        {{\sigma }^{\mu \nu}{e_R}{F_{\mu \nu}}} \nonumber \\
    &&   +{g_1}(\overline{{{\mu }_R}}{e_L})
              (\overline{{e_R}}{e_L})
       + {g_2}(\overline{{{\mu }_L}}{e_R})
              (\overline{{e_L}}{e_R}) \nonumber \\
    &&   +{g_3}(\overline{{{\mu }_R}}{{\gamma }^{\mu }}{e_R})
              (\overline{{e_R}}{{\gamma }_{\mu }}{e_R})
       + {g_4}(\overline{{{\mu }_L}}{{\gamma }^{\mu }}{e_L})
              (\overline{{e_L}}{{\gamma }_{\mu }}{e_L})  \nonumber \\
    &&   +{g_5}(\overline{{{\mu }_R}}{{\gamma }^{\mu }}{e_R})
              (\overline{{e_L}}{{\gamma }_{\mu }}{e_L})
       + {g_6}(\overline{{{\mu }_L}}{{\gamma }^{\mu }}{e_L})
              (\overline{{e_R}}{{\gamma }_{\mu }}{e_R})
       +  h.c. \},
\end{eqnarray}
where $A_{R}$ and $A_{L}$ are the photon-penguin
 type coupling constants which contribute to
 both the $\mu \rightarrow e \gamma$ and $\mu \rightarrow 3e$ 
processes and $g_{1-6}$ are 
the four-fermion type coupling constants which contribute to 
 only the $\mu \rightarrow 3e$ process.
 The Fermi constant $G_F$ and the muon mass $m_{\mu}$ is
 factored out and these
 coupling constants are dimensionless.
These effective coupling constants have a specific pattern
 for the two SUSY GUTs.
In the SU(5) case, only the right-handed sleptons have
 LFV coupling constants.
Then, $A_R$ is suppressed by $m_e/m_{\mu}$ relative to $A_L$
 because a chirality flip must occur at the electron side.
Among the four-fermion type coupling constants,
$g_1$ and $g_2$ are suppressed relative to the other coupling constants
 by the Yukawa coupling constant because they need a chirality flip.
Then, $g_3$ and $g_5$ dominate the process.
On the other hand, in the SO(10) case, both the left-handed
 and right-handed sleptons have LFV coupling constants.
As a consequence, photon-penguin type diagrams which pick up
 the tau mass as a chirality flip become possible because
 the slepton in the loop diagram can change its flavor before
 and after the chirality flip\cite{bhs95}.
They are enhanced by $m_{\tau}/m_{\mu}$ compared to the other diagrams.
Then, $A_R$ and $A_L$ dominate the $\mu \rightarrow 3e$ process.

\section{P- and T-odd asymmetries in $\mu \rightarrow e \gamma$
 and $\mu \rightarrow 3e$ processes}
Now that we explained the qualitative features of LFV effective coupling constants
in the SU(5) and SO(10) SUSY GUT,
let us define $\lambda_{\tau}$ 
insensitive observables: the P- and T-odd asymmetries. 
In the case of the $\mu^+ \rightarrow e^+ \gamma$ process,
 the differential branching ratio is written as follows:
\begin{eqnarray}
\frac{dB(\mu \rightarrow e \gamma)}{d\cos\theta}
             &=& \frac{B(\mu \rightarrow e \gamma)}{2}
                 \{1+A(\mu \rightarrow e \gamma)P\cos\theta\},\\
A(\mu \rightarrow e\gamma) &=& \frac{|A_L|^2-|A_R|^2}{|A_L|^2+|A_R|^2},
\end{eqnarray}
where $\theta$ is an angle between the decay positron momentum and
 the initial muon polarization. 
%
We define the coefficient of the angle dependence as the 
P-odd asymmetry $A(\mu \rightarrow e \gamma)$.
 In the case of the $\mu^+ \rightarrow e^+e^+e^-$ process,
the kinematics is determined by energies of two decay
positrons $E_1$, $E_2$ ($E_{1} > E_{2}$) and two angle which 
indicate the direction of muon polarization to the decay 
plane $\Omega(\theta,\phi)$ (See Figure \ref{fig:su5} (a)).
In the coordinate of Figure \ref{fig:su5} (a), components of
 the initial muon polarization are written using the momenta of the
 decay positron and electron as follows:
\begin{eqnarray}
\vec{P} &=& (~\vec{P} \cdot \frac{\vec{p_1}-(\vec{p_1} \cdot \frac{\vec{p_3}}{|p_3|})
\frac{\vec{p_3}}{|p_3|}}{|\vec{p_1}-(\vec{p_1} \cdot \frac{\vec{p_3}}{|p_3|})
\frac{\vec{p_3}}{|p_3|}|},~
\vec{P}\cdot \frac{\vec{p_3}\times\vec{p_1}}{|\vec{p_3}\times\vec{p_1}|},~
\vec{P}\cdot\frac{\vec{p_3}}{|\vec{p_3}|}~).
\end{eqnarray}
The x and z components of the polarization are P-odd 
because the momenta change their sign under the parity (P) 
operation and the polarization does not.
The y component of the polarization is T-odd 
because the momenta and the polarization change
their sign under the time reversal (T) operation.
Then three asymmetries of branching ratio can be defined relative to the 
muon polarization as follows:
\begin{eqnarray}
A_i &=& \frac{\int_{P_i>0} d\Omega \frac{dB(\mu\rightarrow 3e)}{d\Omega}
                  -\int_{P_i<0} d\Omega \frac{dB(\mu\rightarrow 3e)}{d\Omega}}
             {B(\mu\rightarrow 3e)},~~~~~(i=x,y,z).
\end{eqnarray}
We call the asymmetry relative to the z component as $A_{P_1}$,
 that of the x component as $A_{P_2}$ and
 that of the y component as $A_T$.
These asymmetries can be written in terms of 
 combinations of the effective coupling 
constants.
The P-odd asymmetries $A_{P_1}$ and $A_{P_2}$ reflect
 a chiral structure of these coupling constants.
The T-odd asymmetries can be induced only through
 the interference between the photon-penguin and four-fermion
 type coupling constants as follows:
\begin{eqnarray}
A_{T} &\equiv& A_y \simeq \frac{3
                            [2Im(eA_Rg_4^*+eA_Lg_3^*)
                          -1.6Im(eA_Rg_6^*+eA_Lg_5^*)]}{2B(\mu\rightarrow 3e)}, 
\end{eqnarray}
where we set a cut-off in the positron energy ($E_{1,2}<0.04 m_{\mu}$) 
 because near the kinematical edge ($E_{1,2}\simeq \frac{m_{\mu}}{2}$)
 the contribution to the denominator from the photon-penguin
 amplitudes has a logarithmic singularity and dilutes
 $A_T$.

\section{Results of numerical calculation}
\begin{figure}[t]
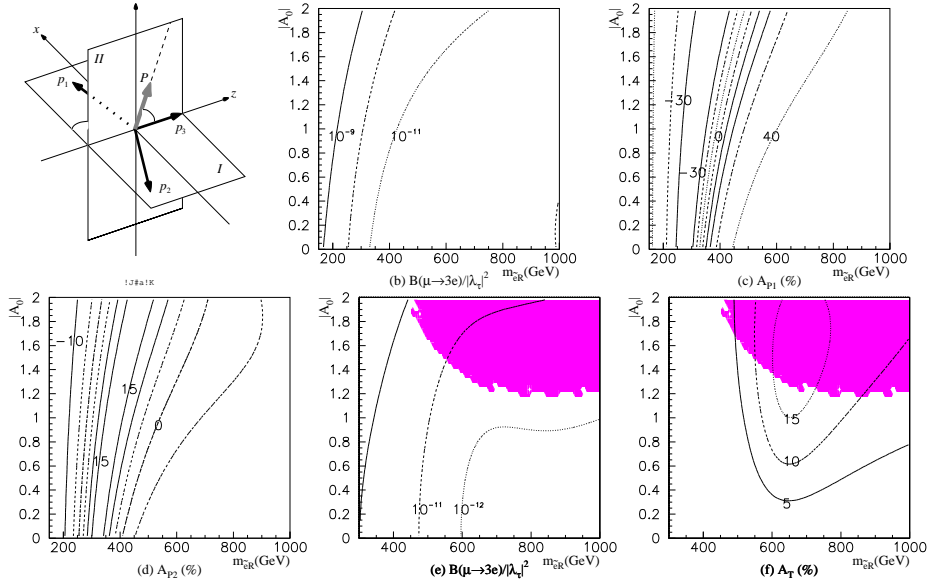

~\psfig{figure=fig1.epsi,height=1.5in}~
\psfig{figure=fig3b.epsi,height=1.5in}~
\psfig{figure=fig3e.epsi,height=1.5in}
~\psfig{figure=fig3f.epsi,height=1.5in}~
\psfig{figure=fig4b.epsi,height=1.5in}~
\psfig{figure=fig4g.epsi,height=1.5in}
\caption{Branching ratio and P- and T-odd asymmetries of $\mu^+ \rightarrow
 e^+e^+e^-$ in the SU(5) SUSY GUT for $\tan\beta=3$. The SUSY parameters are
 fixed as $M_0=150$ GeV and $\theta_{A_0}=\theta{\mu}=0$ for (b)-(c)
 and $M_0=300$ GeV, $\theta_{A_0}=\frac{\pi}{2}$ and $\theta_{\mu}=0$ for (e), (f). 
The black shaded region is excluded by the experimental bounds for the neutron,
 electron and Hg EDMs.\label{fig:su5}}
\end{figure}

\begin{figure}[t]
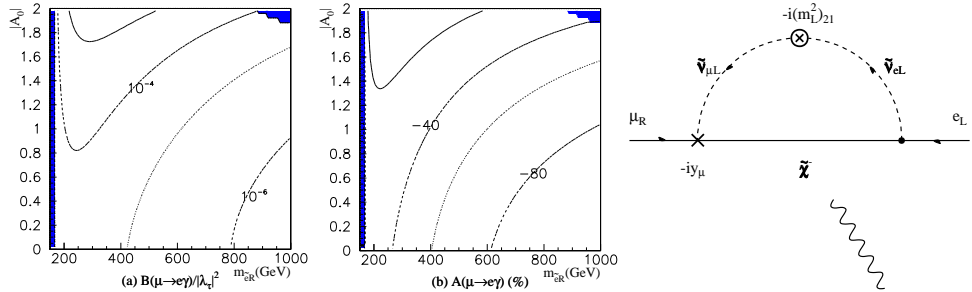

~\psfig{figure=fig6a.epsi,height=1.5in}~
\psfig{figure=fig6d.epsi,height=1.5in}~
\psfig{figure=mueg3.epsi,height=1.5in}
\caption{Branching ratio and P-odd asymmetry of $\mu^+ \rightarrow e^+\gamma$ in the SO(10) SUSY GUT for $\tan\beta=3$. The SUSY parameters are fixed as $M_0 = 150$ GeV and $\theta_{A_0}=\theta_{\mu}=0$.\label{fig:so10}}
\end{figure}

 Now, let us show the results of our 
numerical calculations for these asymmetries.
 In the actual numerical analysis, we solved the renormalization group equations 
 from the Planck scale to the EW scale
with full flavor mixings.
 We assumed the minimal-SUGRA-type universal boundary
 conditions for the soft SUSY breaking parameters at
 the Planck scale.
All the scalar fields have a mass $m_0$ and the gaugino has a mass $M_0$. 
The trilinear scalar coupling constants are proportional to
 the Yukawa coupling constants with a universal coefficient $m_0 A_0$.
 In general, these parameters can have 
complex phases and two of them are physically independent.
We take the phase of $A_0$ ($\theta_{A_0}$)
 and the phase of the $\mu$ term ($\theta_{\mu}$) in the MSSM superpotential
  as SUSY CP violating phases.
 We also assumed the radiative EW symmetry breaking.
 Then, the free parameters of the two SUSY GUTs are 
$m_{0}$, $M_{0}$, $|A_{0}|$, $\theta_{A_{0}}$, 
$\theta_{\mu}$ and the ratio of two VEVs of the Higgs
 doublets $\tan\beta$.
The SUSY CP violating phases induce the electron, 
neutron and atomic EDMs through one loop diagrams.
We included such EDM constraints.
 Phenomenological constraints from LEP, Tevatron and 
$b \rightarrow s\gamma$ decay are also imposed.
 First we show the results of the SU(5) SUSY GUT with no SUSY 
CP violating phase in Figure \ref{fig:su5} (b)-(d). 
In this model, $A_{L}$, $g_{3}$ and $g_{5}$ give sizable contributions.
SUSY parameters are fixed as $\tan\beta=3$ and 
 $M_{2}=150$ GeV.
%
The ratio of the two branching fractions $B(\mu \rightarrow 3e)/B(\mu \rightarrow e\gamma)$
is known to become constant ($\simeq 0.0061$) when the photon-penguin diagram dominates
the $\mu \rightarrow 3e$ process.
In our numerical calculation, it is enhanced in a wide region of parameter space
 compared to the photon-penguin dominant case because of the contribution of the 
four-fermion type coupling constants.
In a large parameter space, the ratio is enhanced by more than five.
Figure \ref{fig:su5} (b) show the branching ratio for $\mu \rightarrow 3e$
 normalized by $|\lambda_{\tau}|^2$.
 If we take $V_R$ as corresponding CKM matrix elements, $|\lambda_{\tau}|^2$
 is approximately $10^{-7}$, however, if $\lambda_{\tau}$ is enhanced by a 
 factor of 10 due to higher dimensional operators for the Yukawa coupling constant
 at the GUT scale, the branching ratio becomes $10^2$ times larger. 
 We checked the P-odd asymmetry of $\mu \rightarrow e 
\gamma$ becomes almost $100\%$ as expected.
Figure \ref{fig:su5} (c), (d) show the P-odd asymmetries $A_{P_1}$ and $A_{P_2}$
 of the $\mu \rightarrow 3e$ process.
The $A_{P_1}$ varies from $-30\%$ to $40\%$ and 
 $A_{P_2}$ varies from $-10\%$ to $15\%$
 parameter sensitively.
It is interesting that the effective coupling constants
$g_{3}$, $g_{5}$ and $A_{L}$ can be determined  up to the overall phase
from $B(\mu \rightarrow e \gamma)$, $B(\mu \rightarrow 3e)$ and 
$A_{P_1}$.
Then, $A_{P_2}$ is predicted and can be used for 
checking the assumption.
Next, we introduce the SUSY CP violating phases and 
 show the contour plot of the T-odd
 asymmetry in Figure \ref{fig:su5} (f).
The SUSY parameters are chosen as $\tan\beta=3$, 
$M_{2}=300$ GeV, $\theta_{A_{0}}=\pi/2$.
$\theta_{\mu}$ is constrained very severely by the EDM experiments
 and we consider a small allowed region around $\theta_{\mu}=0$. 
Figure \ref{fig:su5} (e) also shows the branching ratio of $\mu \rightarrow 3e$ 
 divided by $|\lambda_{\tau}|$ for corresponding SUSY parameters.
The black shaded region indicates excluded region 
 by the current experimental bounds for the electron, neutron and Hg EDMs.
Within the EDM constraints, it is shown that the T-odd asymmetry can
 become $15\%$.
 Next, we show the results of the SO(10) SUSY GUT.
In this case, the photon-penguin type coupling constants
$A_{L}$ and $A_{R}$ dominate the $\mu \rightarrow 3e$ process.
Figure \ref{fig:so10} (a) shows $B(\mu \rightarrow e \gamma)$ divided
 by $|\lambda_{\tau}|^2$.
The SUSY parameters are fixed as $\tan\beta=3, 
M_{2}=150$ GeV, $\theta_{A_{0}}=0$ and $\theta_{\mu}=0$.
We confirmed the ratio of two branching fraction is almost
constant as expected.
Figure \ref{fig:so10} (b) shows the P-odd asymmetry of
 $\mu \rightarrow e \gamma$.
The asymmetry varies from $-20\%$ to $-90\%$ and 
the absolute value becomes larger with the mass of 
right-handed selectron.
This result can not be explained if only the diagrams enhanced
 by $m_{\tau}$ dominate the process as believed 
previously.
In such a case $A_{L}$ and $A_{R}$ have the same contribution. 
Instead we found that the chargino 
contribution which only contributes to $A_{R}$ 
can not be neglected in spite of no $m_{\tau}$ 
enhancement (See Figure \ref{fig:so10} (c)).
The chargino contribution even dominates the neutralino
 contribution enhanced by $m_{\tau}$ when the scalar mass becomes large.
This is mainly
because the dominant diagram of the former contribution picks up a chirality flip
 with a factor $m_{\mu}/m_W$ at the Higgsino vertex whereas the
 latter picks up it by the left-right mixing mass in the slepton internal line
 with a factor $m_{\tau}/m_{\tilde{\tau}}$ which decreases with the scalar mass. 
The P-odd asymmetries $A_{P_1}$ and $A_{P_2}$  are simply
 proportional to $A(\mu \rightarrow e \gamma)$ because in the SO(10) GUT
 only the photon-penguin type coupling constants 
dominate the process and there is essentially only one 
P-odd asymmetry.
The T-odd asymmetry is found small because the T-odd 
asymmetry occurs only through the interference terms of
the photon-penguin and four-fermion type coupling constants and 
it is suppressed when the photon-penguin contributions 
dominate the process.
Table \ref{tbl:summary} is the summary of our numerical calculation.
These results make a good contrast between the two SUSY GUTs.
In this article we discussed the possibility to use various P- and T- odd
 asymmetries relative to the muon polarization 
in the $\mu \rightarrow e \gamma$ and $\mu \rightarrow 3e$
to explore LFV interactions at the GUT scale. 
As a result of a detailed numerical calculation,
we showed these observables and the ratio of two branching fractions are
 useful to extract the difference of the SU(5) and SO(10) SUSY GUT.
\begin{table}[t]
\caption{Summary of the numerical calculation\label{tbl:summary}}
\vspace{0.2cm}
\begin{center}
\footnotesize
\begin{tabular}{|c|c|c|} 
\hline
 & SU(5) SUSY GUT & SO(10) SUSY GUT \\
\hline
$\frac{B(\mu \rightarrow 3e)}{B(\mu \rightarrow e \gamma)}$
 & $0.007$~--~$O(1)$
 & constant~($\sim 0.0062$) \\
\hline
$A(\mu \rightarrow e \gamma)$ & $+100\%$ & $+100\%$~--~$-100\%$ \\
\hline
$A_{P_1}$
 & $-30\%$~--~$+40\%$
 & $A_{P_1}\simeq-\frac{1}{10}A(\mu \rightarrow e \gamma)$ \\
\hline
$A_{P_2}$
 & $-20\%$~--~$+20\%$
 & $A_{P_2}\simeq -\frac{1}{6}A(\mu \rightarrow e \gamma)$ \\
\hline
$|A_T|$
 & $\lsim 15\%$
 & $\lsim 0.01\%$ \\ 
\hline
\end{tabular}
\end{center}
\end{table}

\end{document}